\documentclass[11pt]{article}
\usepackage{moriond,epsfig}
\usepackage{graphicx,amsmath,amssymb,epsf, fdag, bm}

\def\be{\begin{equation}}
\def\ee{\end{equation}}
\def\bea{\begin{eqnarray}}
\def\eea{\end{eqnarray}}

\begin{document}
\vspace*{4cm}
\title{Unintegrated sea quark at small x and vector boson production}
\author{F.\ Hautmann} 
\address{University of Oxford,    Oxford OX1 3NP}

\author{ Martin Hentschinski}

\address{Instituto de F{\' i}sica Te{\' o}rica  
UAM/CSIC,    Universidad Aut{\' o}noma de Madrid, E-28049 Madrid \\
Physics Department, Brookhaven National Laboratory, Upton NY 11973   }

\author{H.\ Jung}
\address{Deutsches Elektronen Synchrotron, D-22603 Hamburg  }

\maketitle

\abstracts{We discuss recent work on the transverse momentum dependent sea quark 
density and its application to forward Drell-Yan production.}

\vspace{-12cm}
\begin{flushright}
  IFT-UAM/CSIC-12-45
\end{flushright}
\vspace{10.5cm}
\section{Introduction}\label{sec:intro}

Scattering processes with a single hard scale are well described   in QCD 
within the framework of collinear factorization. The treatment of
multi-scale processes,  on the other hand, is  more involved.  In this
case, generalized factorization formulas  are needed    \cite{jcc-book} 
to gain control
over large logarithms in higher orders of perturbation theory.  
Such  formulas typically   
involve transverse-momentum dependent (TMD), or ``unintegrated'',  
parton distribution and parton decay functions.
A broad class of  multiple-scale events is given by small-$x$
processes.  These  are one of the main sources of final states in
the central region at the LHC,   and lead to sizeable  
rates of forward  jet production at the LHC
\cite{fwdjetphys1,fwdjetphys2}.  At small $x$, TMD parton  
distributions arise naturally as a consequence of high energy   
factorization and BFKL evolution \cite{Fadin:1975cb}. $k_T$-factorization   
\cite{ktfac,Catani:1994sq}  provides then the
matching of these high energy factorized TMD distributions to
collinear factorized distributions.  For Monte Carlo applications a
convenient description is given in terms of the CCFM evolution
equation \cite{Ciafaloni:1987ur} which 
interpolates for inclusive observables between DGLAP and BFKL
evolution \cite{hj_rec}. 
This  therefore supplies a natural basis for a Monte-Carlo  
realization of $k_T$-factorization,  such as that  provided by the Monte  
Carlo event generator \textsc{Cascade}~\cite{casc}.

Computational tools based on TMD parton densities have so far been
developed within a quenched approximation where only gluons and
valence quarks are taken into account~\cite{fwdjetphys2,hj_ang}.
While this captures correctly the leading contributions at small $x$,
it is mandatory to go beyond this approximation in order to include
preasymptotic effects and to treat final states associated with
quark-initiated processes such as Drell-Yan production.

In this contribution we present work~\cite{Hautmann:2012sh} 
  in this direction,  and its application to  forward Drell-Yan
production. For further  detail   we refer to~\cite{Hautmann:2012sh,hhj_pho}.

\section{Definition of a TMD sea quark distribution and off-shell $qq^* \to Z $ coefficient}
\label{sec:defsea}

The   unintegrated
sea-quark distribution  is  analyzed in~\cite{Hautmann:2012sh}  
to logarithmic accuracy  $\alpha_s (\alpha_s \ln x)^n $ 
 based on  the off-shell TMD
gluon-to-quark splitting function~\cite{Catani:1994sq}.
This  is obtained by generalizing the expansion in two-particle
irreducible kernels of~\cite{CFP} to finite transverse
momenta, and  reads 
\begin{align}
  \label{eq:ktsplitt_def}
  {P}_{qg} \left(z, \frac{{\bm k}^2}{{\bm \Delta}^2}\right) = 
T_R \left( 
            \frac{{\bm \Delta}^2}{{\bm \Delta}^2 + z(1-z){\bm k}^2}
\right)^2 
\left[
{(1-z)^2 + z^2 }
 + 4z^2 (1-z)^2 \frac{{\bm k}^2}{{\bm \Delta}^2}
\right].
\end{align}
Here ${\bm \Delta} = {\bm q} - z\cdot {\bm k}$ with ${\bm k}$ and ${\bm q}$
transverse momenta of the off-shell gluon and quark respectively,
while $z$ is the fraction of the  `minus' light cone momentum of the gluon which is 
carried on by the $t$-channel quark.
 Although  evaluated off-shell, the splitting probability
 is universal.   Once 
combined  with the gluon Green's function, it takes into account the 
 small $x$ enhanced transverse momentum dependence to all orders 
in the strong coupling. In this 
approach the transverse momentum  of the sea quark  arises  
 as a consequence of subsequent  branchings at small $x$, 
 with  no  strong ordering in their transverse momenta.   
\begin{figure}[h]
  \centering
  \parbox{4cm}{\includegraphics[width = 3.8cm]{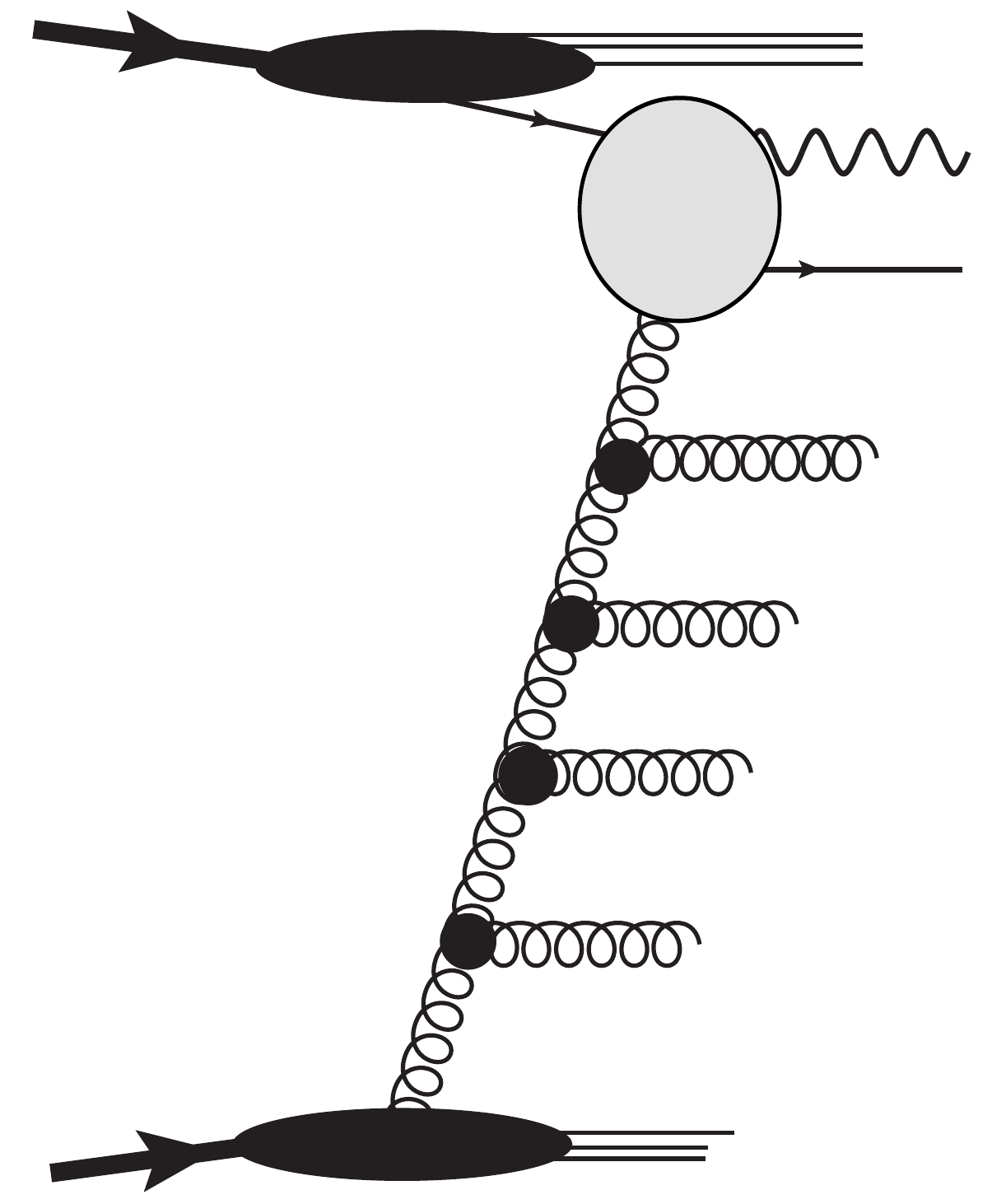}} \parbox{2cm}{$\!$}    \parbox{2.5cm}{\includegraphics[height = 2.3cm]{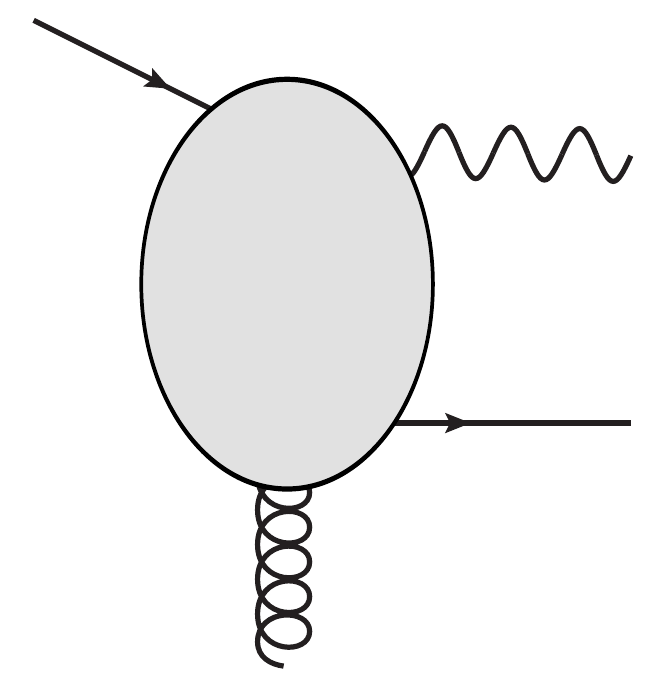}}  \parbox{1.4cm}{\includegraphics[width = .7cm]{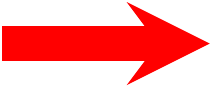}}
\parbox{1.8cm}{\includegraphics[height = 1.7cm]{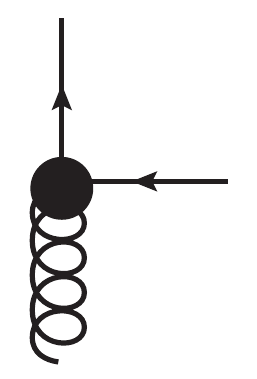}} ${ \bm  \otimes}$  \parbox{2cm}{\includegraphics[height = 1.5cm]{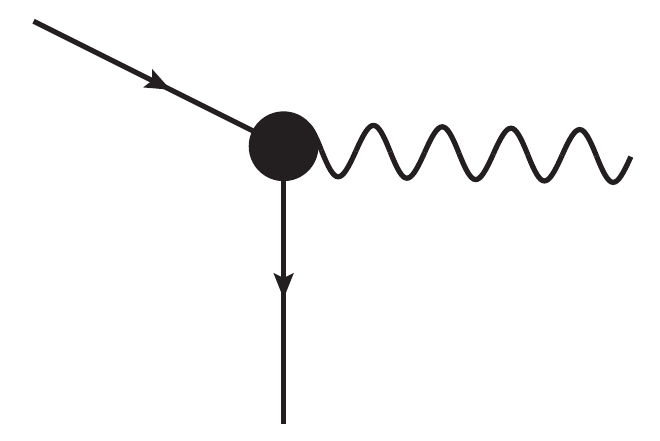}} 

\parbox{4cm}{\center (a)} \parbox{9.7cm}{\center (b)}

\caption{
      \small (a): If the vector boson is produced in the forward
  region, the sea quark density becomes sensitive to multiple small $x$
  enhanced gluon emissions, leading to a $k_T$-dependent gluon density
  (b): Schematic factorization of the partonic $qg^* \to Zq$ process of
  a) into  the $g^* \to q^*$ splitting and the $qq^* \to Z$ coefficient.
  }
  \label{fig:splitting}
\end{figure}

To relate this parton splitting kernel to forward vector boson
production, we analyze the flavor exchange process $g^*q \to Zq $, see
 Fig.~\ref{fig:splitting}. At high (partonic) center of mass
energy, this process can be treated according to the ``reggeized
quark'' calculus~\cite{Lipatov:2000se,Bogdan:2006af}.  The latter
extends the effective action formalism~\cite{Lipatov:1995pn},
currently explored at NLO~\cite{Hentschinski:2011tz}, to amplitudes
with quark exchange in terms of effective degrees of freedom, the
so-called reggeized quarks~\cite{Fadin:1976nw,knie-sale}.  The use of
the effective vertices~\cite{Lipatov:2000se,Bogdan:2006af} ensures
gauge invariance of the coefficients relevant to perform the
high-energy factorization~\cite{ktfac,Catani:1994sq} for vector boson
production, despite the off-shell parton.  

If taken literally, the
reggeized quark calculus leads for the $g^*q \to Zq$ process to a
rather crude approximation to the $g^* \to q^*$ splitting function, 
 associated with the  lightcone momentum ordering condition which sets the `plus'
momenta of the off-shell quark for the $g^* \to q^*$ splitting to
zero. For Eq.~(\ref{eq:ktsplitt_def}) this corresponds to the limit $z
 \to  0$. It is however possible~\cite{Hautmann:2012sh} 
  to relax this kinematic restriction and
to keep $z$ finite, while maintaining the gauge invariance properties
of the original vertex.  For the $g^* \to q^*$ splitting this yields
then precisely the splitting function Eq.~(\ref{eq:ktsplitt_def}).

On the other hand,  in the $qq^* \to Z$ coefficient  the high energy limit sets the
`minus' component of the quark momentum to zero.   It proves to be  possible to
relax the ordering prescription also in this case.  
It is thus interesting to investigate the effect  of these kinematic corrections, which are 
subleading in the collinear   and high energy limits. 
In~\cite{Hautmann:2012sh} we express the off-shell coefficient for the 
$Z$-boson cross section as 
\begin{align}
\label{eq:part_tfac}
  \hat{\sigma}_{qq^* \to Z}
&=
\sqrt{2} G_F M_Z^2 (V_q^2 + A_q^2)  \frac{\pi}{N_c} \delta(z x_1 x_2 s  + T - M_Z^2).
\end{align}
Here the variable $T$ parametrizes the off-shellness of the
$t$-channel quark. In the collinear limit  $T \to  0$  so that 
 Eq.~\eqref{eq:part_tfac} agrees with  the  lowest order  $qq \to Z$ coefficient. 
For the general off-shell case,  $T$  interpolates between the 
 squared transverse momentum of the off-shell quark,  if  strong minus momentum ordering is fulfilled,  
 and    modulus of  the four-momentum transfer, if this condition is  relaxed.   
 Correspondingly, the  
$qg^* \to qZ$ cross section is 
expressed in terms of   convolutions in  transverse  momentum 
and four momentum transfer~\cite{Hautmann:2012sh}    respectively.

\section{Numerical analysis}
\label{sec:numeric}

\begin{figure}[h!]
  \centering
  \parbox{.49 \textwidth }{\includegraphics[width = .45 \textwidth ]{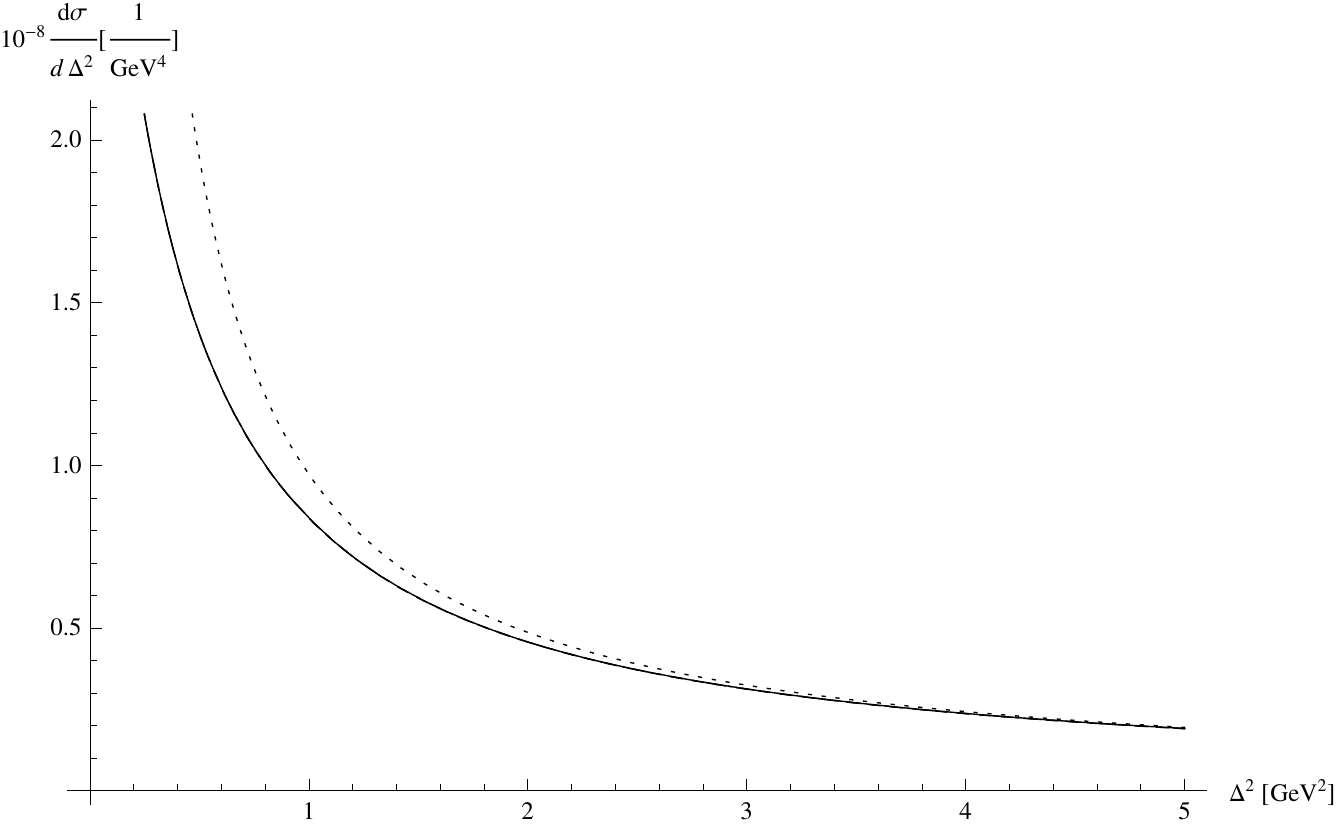}}  
\parbox{.49 \textwidth }{\includegraphics[width = .45 \textwidth ]{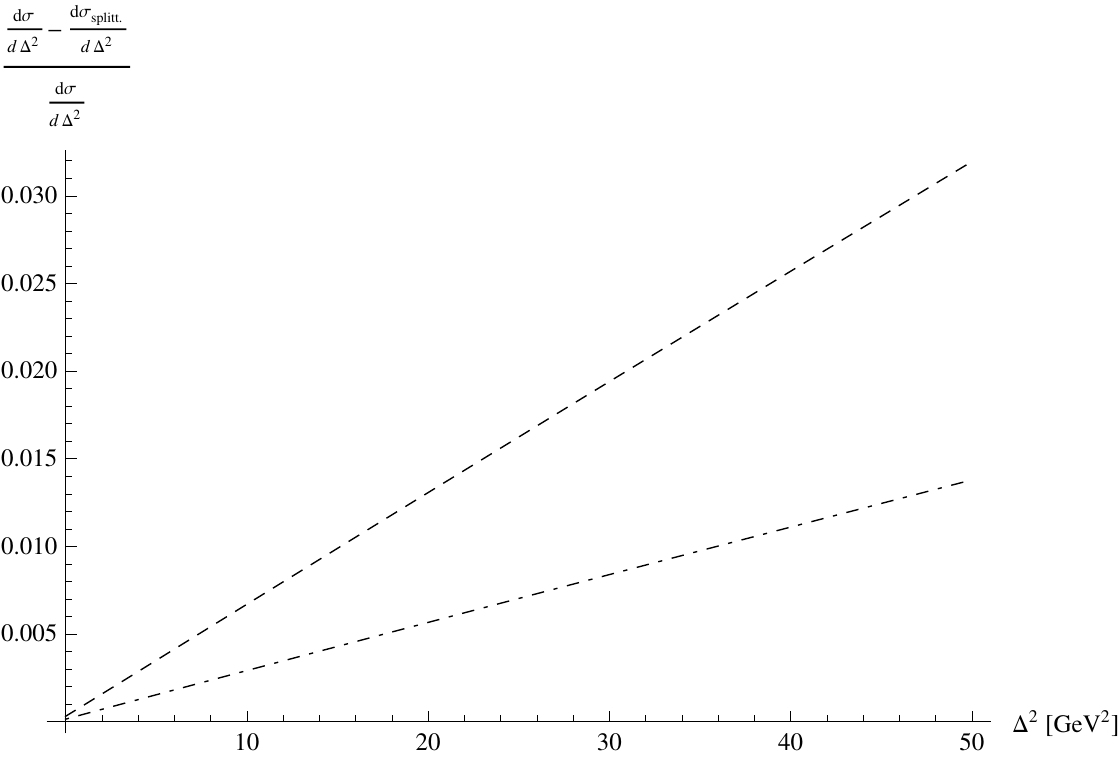}}   \parbox{.49 \textwidth}{\center (a)}  
\parbox{.5 \textwidth}{\center (b)} \\
 \caption{(a):    ${\bm \Delta}^2 $  dependence of    the   
   differential cross section  $ d\sigma /  d {\bm \Delta}^2 $ for   small $| {\bm \Delta} | $: (solid) full; (dashed) no plus-momentum ordering; (dot-dashed) no plus-momentum and minus-momentum ordering;  (dotted)  collinear approximation.  All but the last curve overlap in this region.  We set 
 $x_1 x_2 s = 2.5 M_Z^2 $, ${\bm k}^2 = 2$ GeV$^2$.  (b):~Relative deviations  in   
 the differential cross section $ d\sigma /  d {\bm \Delta}^2 $:  (dashed) no plus-momentum ordering; 
 (dot-dashed)    no plus-momentum and minus-momentum ordering.}
 \label{fig:deltasmall}
\end{figure}

Fig.~\ref{fig:deltasmall} shows a 
 numerical comparison~\cite{Hautmann:2012sh} 
       of the  factorized  formulas discussed above 
   with the 
  $qg^* \to qZ$ matrix element result and 
 with  an expression which uses only the collinear splitting function.
  For small $| {\bm \Delta} | $,  the differences between $t$ and $k_T$-factorized expressions
are numerically small, and  both expressions are close to the full
result; as $| {\bm \Delta} | $ increases, we find that  the deviations
due to the kinematic contributions by which the two expressions   differ
become non-negligible, and that the $t$-factorized expression gives a
better approximation to the full result.

Future extensions of the above results concern 
large-$x$ contributions~\cite{idi-sci,ceccopie,chered,jain-proc,jain,jccfh01};  
parton shower Monte Carlo 
implementations~\cite{hhj_pho};  
 inclusion of full quark emissions in  the evolution, 
 see~\cite{Hentschinski:2011ft} for related work in the context of  
  next-to-leading order BFKL evolution.

\section*{Acknowledgments}
We thank the organizers for the opportunity to present this work at the meeting. M.~H.\  is grateful for financial support from the German Academic   
Exchange Service (DAAD), the MICINN under grant FPA2010-17747, the
Research Executive Agency (REA) of the European Union under the Grant
Agreement number PITN-GA-2010-264564 (LHCPhenoNet), the Helmholtz
Terascale Analysis Center, the U.S. Department of Energy under contract number DE-AC02-98CH10886 and a BNL “Laboratory Directed Research and Development” grant (LDRD 12-034).

\end{document}